\documentclass[a4paper,twocolumn,11pt,accepted=2025-06-03]{quantumarticle}
\pdfoutput=1
\usepackage[utf8]{inputenc}
\usepackage[english]{babel}
\usepackage[T1]{fontenc}
\usepackage{amsmath}
\usepackage{hyperref}

\usepackage{tikz}
\usepackage{lipsum}

\usepackage[numbers,sort&compress]{natbib}

\begin{document}

\title{Generalizing the matching decoder for the Chamon code}

\author{Zohar Schwartzman-Nowik}
\affiliation{School of Computer Science and Engineering, Hebrew University, Jerusalem, Israel}
\affiliation{Faculty of Engineering and the Institute of Nanotechnology and Advanced Materials, Bar Ilan University, Ramat Gan, Israel}
\orcid{0000-0003-3804-4933}
\email{zohar.nowik@mail.huji.ac.il}
\author{Benjamin J. Brown}
\affiliation{IBM Denmark, Sundkrogsgade 11, 2100 Copenhagen, Denmark}

\maketitle

\begin{abstract}
    Different choices of quantum error-correcting codes can reduce the demands on the physical hardware needed to build a quantum computer. To achieve the full potential of a code, we must develop practical decoding algorithms that can correct errors that have occurred with high likelihood.
    Matching decoders are very good at correcting local errors while also demonstrating fast run times that can keep pace with physical quantum devices.
    We implement variations of a matching decoder for a three-dimensional, non-CSS, low-density parity check code known as the Chamon code, 
    which has a non-trivial structure that does not lend itself readily to this type of decoding.
    The non-trivial structure of the syndrome of this code means that we can supplement the decoder with additional steps to improve the threshold error rate, below which the logical failure rate decreases with increasing code distance.
    We find that a generalized matching decoder that is augmented by a belief-propagation step prior to matching gives a threshold of $ 10.5\%$ for depolarizing noise. 
\end{abstract}

\section{Introduction}

Quantum computers hold great potential to bring significant improvements
in computation compared to their classical counterparts. One of the main
obstacles for realizing these advantages is overcoming the noise obscuring
the information in the quantum devices. We employ quantum error correction codes (QECCs) to overcome this obstacle~\cite{aharonov1997fault, dennis2002topological}. A QECC is a system where
the information is encoded in a subspace of a larger Hilbert space, called
the code space. We operate QECCs by measuring stabilizer checks to obtain syndrome data. This data is then used to determine the best way to correct the errors the system experiences. 
The QECCs we need to build a scalable quantum computer have a high resource cost in terms of physical hardware. The resource cost of building a quantum computer is a function of our choice of QECCs that underlie our quantum computer, and how these QECCs are operated. Finding better ways of operating new QECCs with good parameters can reduce the resource cost of a large-scale quantum computer, and bring their realization forward.

A decoder is an algorithm that takes syndrome data and decides what recovery operation is most likely to restore the logical encoded state to the state that was encoded initially. Decoding is a highly non-trivial task, and a number of  different methods of decoding have been proposed and tested~\cite{demarti2024decoding,  DuclosCianci2010,  bravyi2011analytic, Wootton2012, delfosse2021almost, bravyi2014efficient} that trade-off between efficiency and the likelihood that an error is corrected successfully. The minimum-weight perfect-matching (MWPM) decoder has demonstrated both good accuracy and fast decoding times \cite{edmonds1965paths, dennis2002topological, Fowler2012, bonilla2021xzzx, brown2023conservation, higgott2022pymatching, higgott2023sparseblossomcorrectingmillion, Krinner2022, Sundaresan2023, wu2023fusionblossomfastmwpm, jones2024improvedaccuracydecodingsurface, acharya2024quantumerrorcorrectionsurface}, where decoding can be completed by pairing individual syndrome defects in the case of the surface code~\cite{KITAEV20032, bravyi1998quantumcodeslatticeboundary, dennis2002topological}.

In general, the syndrome data of a QECC may give rise to a structure that does not readily lend itself to decoding with a MWPM decoder. However, due to its demonstrated performance when used with specific codes~\cite{dennis2002topological, Wang2010, Fowler2012, Nixon2021, Miguel2023cellularautomaton, Delfosse2014, brown2020parallelized, Tuckett2020,  bonilla2021xzzx, Srivastava2022xyzhexagonal, Sahay2022,  Kubica2023efficientcolorcode, Huang2023, brown2023conservation, benhemou2023minimisingsurfacecodefailuresusing}, it is valuable to explore ways of using MWPM to decode more exotic quantum low-density parity-check (LDPC) codes, and to interrogate their performance. 
To this end, in this work we develop and analyse a generalization of the MWPM decoder \cite{brown2023conservation} on the Chamon code \cite{chamon2005quantum,bravyi2011topological,zhao2023quantum}. This is a three dimensional, non-CSS LDPC code that gives rise to a non-trivial syndrome when affected by errors.

We design our decoder by exploiting code symmetries~\cite{brown2023conservation, brown2020parallelized}. These are overcomplete subsets of stabilizers of the code such that errors always create an even number of syndrome defects with respect to the stabilizers of the symmetry. 
Specifically, a single qubit Pauli error gives rise to at most two syndrome defects on a symmetry.
This property enables us to employ MWPM subroutines to find a correction by pairing syndrome defects. 
Once the matching step is completed, we use the results of these subroutines to find local clusters of syndrome defects that can be corrected independently with a low-weight correction operator. A local correction is then found using an adaptation of the broom algorithm~\cite{bravyi2011analytic}.

For the Chamon code, the code symmetries consist of stabilizers that lie on two-dimensional planes of the three-dimensional model. 
This structure raises some interesting question towards the endeavor of decoding by matching over symmetries. Given that matching is performed on a vanishingly small subset of all the stabilizers, one might ask how this affects the performance of the decoder. On the one hand, one might expect that decoding performance is improved by concentrating fallible matching subroutines on a small subset of the stabilizers. Indeed, examples have shown high performance decoding algorithms can be designed by choosing codes where a decoder can concentrate on smaller subsets of the entire syndrome~\cite{Delfosse2014, Nickerson2019analysingcorrelated, Sahay2022, Srivastava2022xyzhexagonal, Kubica2023efficientcolorcode, brown2020parallelized, Tuckett2020, bonilla2021xzzx, Huang2023, gidney2023newcircuitsopensource}. On the other hand, the two-dimensional matching subroutines are ignoring a significant majority of the syndrome data, and we might expect to be able to improve performance by passing the global syndrome data to the matching subroutines that operate of isolated planes of the code. This question leads us to investigate variations of the basic matching-based decoder.

The first generalization we consider is a decoder where we begin with an initial greedy correction~\cite{anwar2014fast, smith2023local} before matching. Here we propose a correction if a single flip will remove four defects surrounding a qubit.
We find this decoder improves the threshold of the basic matching decoder from $5\%$ to $6\%$. This decoder suggests an advantage is obtained by using the structure of the syndrome data. In contrast, for the qubit implementation of the toric code a similar greedy step reduces the threshold~\cite{anwar2014fast}.

We also consider a generalization of the decoder where we improve the matching step by first performing belief propagation over the global syndrome~\cite{MacKay2003}. The messages that are passed are used to improve the matching subroutines via belief matching~\cite{Higgott2023improved}. Importantly, this passes information from the three-dimensional syndrome to the planes where matching is conducted. Remarkably, we obtain a threshold of $10.5\%$ with this method. This significantly exceeds that found in earlier work on the Chamon code~\cite{zhao2023quantum} where a threshold of $4.92\%$ is obtained. One can also compare our work to Ref.~\cite{brown2020parallelized} where a three-dimensional QECC with similar properties to the Chamon code is studied.

As a final introductory remark, an interesting feature of the Chamon code is that its non-CSS nature enables our matching decoder to treat all types of Pauli errors equally. In contrast, without some novel types of intervention~\cite{fowler2013optimalcomplexitycorrectioncorrelated, Delfosse2014a, Sahay2022, benhemou2023minimisingsurfacecodefailuresusing}, matching decoders for more conventional CSS codes treat Pauli-X and Pauli-Z errors separately~\cite{dennis2002topological, Sahay2022}. However, it is well understood that such treatment can lead to reduced decoding performance~\cite{DuclosCianci2010, Wootton2012}. As such, we believe the example we consider motivates further work into high-threshold decoding algorithms for other non-CSS LDPC codes, e.g.~\cite{Leverrier2022quantumxyzproduct}.

The remainder of this paper is structured as follows. In Sec. II we present the Chamon code. In Sec. III we define the generalized MWPM decoder for the Chamon code, as well as the two improved versions of the decoder by introducing different initial steps. In Sec. IV we present the results, and in Sec. V we conclude.

\section{The Chamon code}

\subsection{The stabilizer group}

The Chamon code is a stabilizer code. A stabilizer code is defined by a set
of commuting Pauli operators $\left\{ S_{i}\right\} $, \cite{gottesman1997stabilizer}, with eigenvalues $\pm 1$. 
The stabilizers are any operators in the group spanned by the generators $S_{i}$.
The logical states are all eigenstates of all stabilizers with eigenvalues 1.
We also define logical operators, these are Pauli operators that commute with the elements of the stabilizer group, but are not themselves stabilizer operators. These operators generate rotations among the encoded states of the stabilizer code.

We measure the error syndrome of a stabilizer code to attempt to identify the error that has occurred. The syndrome $s$ for an arbitrary logical state $|\psi\rangle$ is a list of bits $s_{i}$ where the result of measuring stabilizer generator
$S_{i}$ on $|\psi\rangle$ is $\left(-1\right)^{s_{i}}$. Thus, the
syndrome for a logical state will always be a string of zeros. When
a different string is obtained, an error is detected. 
Any value of 1 in the syndrome is referred to as a syndrome defect.

\begin{figure}
\includegraphics[]{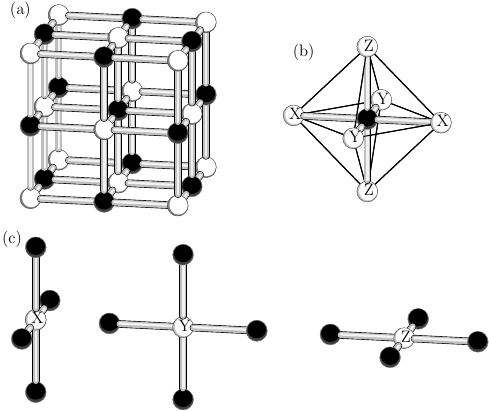}
\caption{\label{fig:chamon code illustaration} The Chamon code. (a) The Chamon code lattice.
Every white vertex represents a qubit and every black vertex represents
a stabilizer. (b) A stabilizer of the Chamon code. All stabilizers in the Chamon code are identical
and apply a Pauli-X operator on the two neighboring qubits in the x axis,
a Pauli-Y operator on the two neighboring qubits in the y axis, and similarly a Pauli-Z
for the z axis. (c) Examples of single-qubit Pauli errors, together with their violated stabilizer operators.}

\end{figure}

We define the stabilizer group of the Chamon code \cite{chamon2005quantum,bravyi2011topological,zhao2023quantum} on a three-dimensional cubic lattice with dimensions $d \times d \times d$ and periodic boundary conditions, where $d$ is the distance of the code, and is even. (A more general definition for non-cubic Chamon codes can be found in \cite{chamon2005quantum, bravyi2011topological}.)
Let us give vertices a coordinate $v = (x,y,z)$ with $1 \le x,y,z \le d$. Vertices are bi-coloured such that white(black) vertices lie on the odd(even) sites where the number $ (x+y+z) $ determines if the vertex $v = (x,y,z)$ is odd(even).
With this bicolouring we have that each white vertex represents a qubit and each black qubit represents
a stabilizer, as can be seen in Fig. \ref{fig:chamon code illustaration}(a). Specifically, for each even vertex $v$ we have a stabilizer
 \begin{equation}
 S_v = X_{v+\hat{x}}X_{v-\hat{x}} Y_{v+\hat{y}}Y_{v-\hat{y}} Z_{v+\hat{z}}Z_{v-\hat{z}} 
\end{equation}
where $\hat{x},\hat{y},\hat{z}$ are unit translations along their respective canonical axes. 
An example of a Chamon code stabilizer is presented in Fig. \ref{fig:chamon code illustaration}(b).
Operators $X_v,Y_v,Z_v$ are standard Pauli matrices acting on qubit $v$ where $v$ is an odd vertex site.

Interestingly, there is a symmetry between the three Paulis - $X$, $Y$ and $Z$ - in the stabilizers of the Chamon code.
Every stabilizer applies exactly two of each Pauli on its neighbors, and every single qubit Pauli error flips the stabilizer measurement of four neighboring stabilizers, as can be seen in Fig. \ref{fig:chamon code illustaration}(c).
Due to this last property, the code is inadequate for a straightforward implementation of the MWPM decoder, and nontrivial adaptations will be needed in order to apply a matching decoder on this code. 

\subsection{The symmetries of the Chamon code \label{symmetries of chamon}}

In this work we propose a matching decoder for the Chamon code. As such, we look for structure in the code that enables us to use matching subroutines to find correctable clusters of the syndrome. To this end we look for code symmetries~\cite{brown2023conservation,brown2020parallelized}. For the Chamon code, we find symmetries of the stabilizer group such that single-qubit Pauli errors always create two syndrome defects with respect to the stabilizers of some given symmetry (or no syndrome defects, if the Pauli error occurred far from the given symmetry). Given that all errors create defects in pairs, we can readily employ matching to group defects together to find correctable clusters.

\begin{figure}
\includegraphics[scale=0.55]{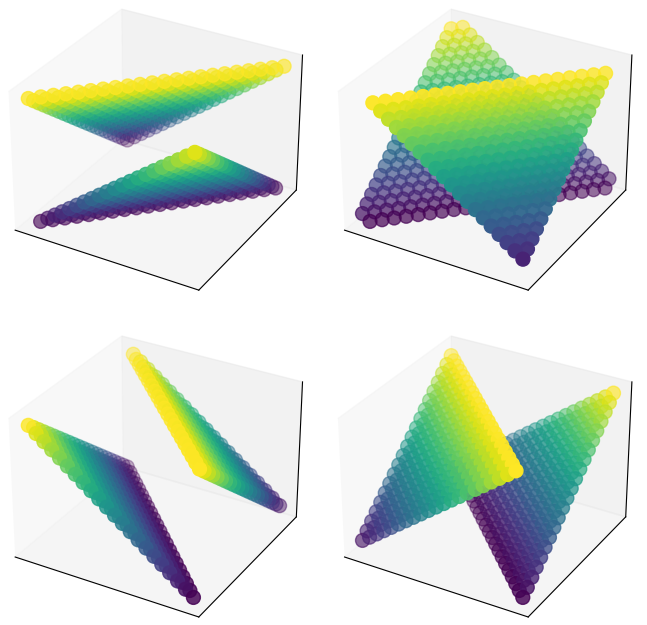}
\caption{\label{fig:chamon sheet example} Examples of the symmetries of the Chamon code, which are the subgroups of stabilizers that are decoded separately. 
In these figures only the stabilizers included in the subgroup are depicted, with the color of the stabilizer indicative of the $z$ value of the stabilizer location. 
In the case of the cubic Chamon
code, these symmetries take the form of
diagonally oriented sheets wrapping once around the periodic boundaries
in each orientation. Representative of the subgroups in each of the four orientations are presented.  }
\end{figure}

The symmetries of the Chamon code that we exploit lie on two-dimensional planes of the three-dimensional code.
We show examples in Fig. \ref{fig:chamon sheet example}. We find the subsets of stabilizers $S_v$ at locations $v = (x,y,z)$ that lie on the planes as follows
\begin{equation}
\Sigma^{\mathbf{r}}(C) = \left\{ S_v  : {(r_x x + r_y y + r_z z)\,(\textrm{mod } d) = \textrm{C}} \right\}, \label{Eqn:Symmetry}
\end{equation}
where the plane lies perpendicular to the vector $\mathbf{r} = (r_x, r_y, r_z)$ with $r_x,r_y,r_z = \pm 1$ and $C$ is a constant that translates the plane along $\mathbf{r}$. Given that $-\mathbf{r}$ is parallel to $\mathbf{r}$, we have four unique plane orientations by this definition. 
There are $d/2$ different symmetries in each orientation, and so a total of $2d$ symmetries.
One can check that for each of these subsets, the product of all the elements of this subset gives identity. It follows from this observation that any Pauli error will always produce an even number of defects among the stabilizers of a code symmetry. In fact, one can check that errors always produce defects in pairs among each of these subsets.

\section{A matching decoder for the Chamon code}

Error correction requires a decoder to attempt to correct the error that has occurred. The decoder uses the syndrome to identify the recovery operation that brings the state back to its original logical state with high probability.
We will assume the noise is Pauli noise for simplicity, as the stabilizer measurements project a general noise channel to a noise close to Pauli noise \cite{bravyi2018correcting, darmawan2017tensor, schwartzman2024modeling}.
For some specific instance of a Pauli error $E$, which specifies which qubits suffered from which single qubit Pauli noise, the stabilizer measurement results in a syndrome $s$.
The decoder takes the syndrome $s$ as input and returns as output a recovery operation $C$ which it deems most probable to bring the state back to its encoded logical state.
When analyzing the logical error rate of a code via simulations where $E$ is known, we can determine whether or not the recovery operation indeed succeeded in bringing the state back to the original logical state by observing whether $CE$ is a logical operator, in which case the decoding failed and resulted in a logical failure, or whether $CE$ is in the stabilizer group, in which case the decoding succeeded.

Here we describe the decoding algorithms that we implement for the Chamon code, with the source code available at \cite{chamon_github}. All the decoders we implement are based on a matching step on symmetries that is followed by a local correction once matching is completed. We describe this first. We then go on to describe modifications to the basic implementation of the matching algorithm. We consider a greedy initialization step, as well as a belief propagation initialization step to improve the accuracy of the MWPM subroutine.

\subsection{Basic matching}

The decoder uses matching subroutines to identify small clusters of syndrome defects that can be corrected locally. The clusters are obtained by connecting defects that share an edge returned from the matching subroutines, where the matching is performed on all planar subsets of stabilizers as defined in Eqn.~(\ref{Eqn:Symmetry}). A pair of defects are included in a common cluster if they share an edge on any matching subroutine. In general, a cluster contains many defects, as each defect will share edges with multiple defects over all of the matching subroutines.

We posit that the clusters obtained by connecting components of the syndrome that share an edge returned from a matching subroutine are locally correctable. It remains then to correct these defect clusters. We find that a correction can be obtained for each of these local clusters using an implementation of the broom algorithm~\cite{bravyi2011analytic} for the Chamon code.
Let us now describe these steps in more detail.

\subsubsection{Step 1 - Matching on symmetries}

\begin{figure}
\includegraphics{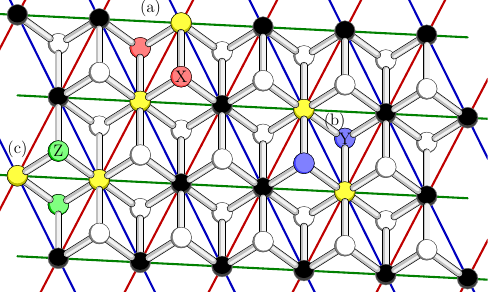}
\caption{Stabilizers included in a planar symmetry, shown in black. Errors violate stabilizers in the plane that create defects. Defects are shown in yellow. A Pauli-X, Pauli-Y and Pauli-Z error are shown at (a),~(b) and~(c), respectively. All examples of errors give rise to pairs of defects. This property enables us to employ matching on individual planes. We find that there are two qubits where errors can give rise to the same pair of defects on the symmetry plane. Indeed, a Pauli-X, Pauli-Y or Pauli-Z error, respectively, on either of the red, blue or green qubits give rise to the same pair of defects on the plane. Different Pauli errors determine the orientation that defects appear on the plane. Pauli-X, Pauli-Y and Pauli-Z errors are always oriented along the red, blue and green lines that underly the lattice, respectively. In the limit that noise is biased such that we only experience one type of Pauli error, all errors give rise to pairs of defects with a common orientation. As such matching on a given plane is reduced to a one-dimensional problem. This is similar to other examples in, e.g., Refs.~\cite{Tuckett2020, bonilla2021xzzx, Huang2023}. In this noise-bias limit the matching decoder we propose for the Chamon code is the same as that presented in Ref.~\cite{Tuckett2020} applied to planes of the code. \label{Fig:SymmetryPlane}}
\end{figure}

On each of the symmetries described in Sec. \ref{symmetries of chamon}, every single-qubit Pauli error
gives rise to up to two syndrome defects, and so can be decoded using
MWPM, see Fig.~\ref{Fig:SymmetryPlane}. The first step of the matching decoder then is to apply matching, implemented using \texttt{PyMatching} \cite{higgott2022pymatching}, on each of the code symmetries, as in Eqn.~(\ref{Eqn:Symmetry}), separately.
Since we are using these decoders as a first step, in a multi-step decoder, we are not interested in the specific qubits it deems should be flipped. Instead, we take as output the previous step in the decoder -- the pairs of syndrome defects it deems should be connected.
And so the output of this first step in the decoding process is a set of connections between the syndrome defects.

\subsubsection{Step 2 - dividing syndrome defects into clusters}

\begin{figure}
\includegraphics[scale=1.0]{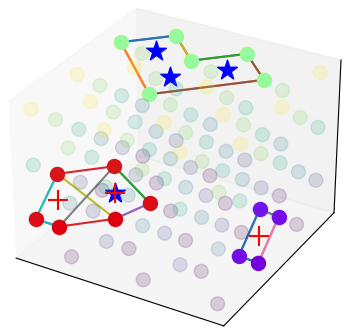}\caption{\label{fig:chamon clusters} An example where syndrome defects are divided into clusters,
together with the connections determined by all the separate matching
decoders on symmetries. Syndrome defects are depicted using non-opaque balls, with different colors determined by the cluster they are included in.
The solid lines between syndrome defects are the connections obtained from MWPM on the different symmetries.
The Pauli errors on
the qubits are also depicted, with a $X$ error depicted as a reg cross,
a $Z$ error as a blue star, and a $Y$ error as a combination of
the two.}
\end{figure}

We can now create a graph of syndrome defects and connections.
The vertices of the graph are the syndrome defects, and the undirected edges are all the connections obtained from all the separate MWPM decoders of the previous step.
The next step of the decoder is to identify all connected components in the graph, in order to later correct each connected component separately. 
This was done by using the \texttt{connected components} function in the \texttt{scipy} package \cite{scipy}.
After identifying the connected components, we divide the vertices of the graph - i.e. the syndrome defects - into the separate sets corresponding to the connected components, which we denote clusters.
An example of such a division in presented in Fig. \ref{fig:chamon clusters}.

The next step will be decoding each cluster of defects separately using a version of the broom decoder, which will be explained in the next step.
In order to apply the broom decoder \cite{bravyi2011analytic}, we need to identify the starting point and orientation to sweep the syndrome defects.
To this end, we must identify the boundaries of a box enclosing each connected component.
This is done by starting with a plane that crosses one of the syndrome defects in the cluster, then lowering the plane until it no longer crosses through a vertex or edge of the connected component.
Repeating this for every axis and every orientation will give us a box enclosing the cluster, if one exists.
If such a box does not exist we can determine with very high probability that the correction will fail.
If we allow discarding samples then the best method would be to discard such a sample, and that will give better success rates.
In this analysis we restrict ourselves by not allowing such post-selection.
So instead, we guess random boundaries for the box, knowing that with very high probability the decoding for these cases will fail.

\subsubsection{Step 3 - correcting each cluster via sweeping decoder}

For low enough error rates, these clusters will not be too large.
Thus, we will decode each cluster by the simple and quick sweeping
decoder, which is an adaptation of the broom decoder \cite{bravyi2011analytic} and unfolds as follows. We pick the directions to sweep,
say x and z directions. We start by sweeping downward in the $z$
direction - we pick a syndrome defect in the top of the cluster and
apply the appropriate single qubit Pauli that swaps this syndrome
defect by three defects in lower layers. We procceed similarly until
all syndrome defects are in a double layer in the bottom of the box.
Next we repeat this proccess in a different orientation, and ``push''
all syndrome defects from front to back. 
After sweeping in two directions the syndrome defects are confined to a very thin tube, and the probability of syndrome defects remaining after this process, for moderate or large system sizes is extremely rare.
In the negligible cases where this occurs, we deem the instance a failure of the decoder.
Otherwise, the recovery process resulting from the decoder are all
single-qubit Paulis that were applied in the sweeping process.

\subsection{Greedy matching}
We suggest an improvement on the basic version of the decoder by adding an initial greedy step to the decoding algorithm. In another context, this greedy step could be interpreted as initialization~\cite{anwar2014fast} or predecoding~\cite{smith2023local}, or an initial stage using a small-set-flip decoder~\cite{Fawzi2018}.

Let us give some intuition for this idea.
A single qubit pauli error gives rise to exactly four syndrome defects, in the shape of a diamond in one of the orientations, as can be seen in Fig. \ref{fig:chamon code illustaration}(c). 
The greedy step we implement consists of a single round over all the syndrome defects where every such diamond pattern is identified.
After all the diamond patterns are identified, the single-qubit Pauli operations that correspond to each of the diamond patterns are added to the recovery operator, eliminating the corresponding syndrome defects.
And so the MWPM on symmetries is then implemented on a reduced number of syndrome defects.

\subsection{Belief matching}

The basic MWPM decoder on symmetries has a major drawback since every MWPM decoder applied on a symmetry has access only to the syndrome defects in that symmetry and no knowledge of the other syndrome defects. 
An improved version of the decoder can overcome this obstacle by adding an initial step of soft belief propagation (BP) decoding \cite{roffe_decoding_2020, Roffe_LDPC_Python_tools_2022}.
Belief propagation decoding assigns initial probabilities for each qubit to suffer from each Pauli error, determined by the probabilities in the noise channel acting on the qubits. Then all qubits send messages with these probabilities to the stabilizers they interact with. The stabilizers then use these messages, together with the information on whether or not they suffer from a syndrome defect, to determine updated probabilities for each Pauli error on each qubit, and send these back as messages. The qubits then infer an updated probability for each Pauli error according to these messages. This is one round of BP.
The output, after multiple rounds of BP, are updates weights assigned to each qubit and each Pauli error corresponding to the probability of such an error occurring on the qubit.
These weights are then used as the initial weights for the matching decoders on each of the symmetries, thus incorporating global information in them.
The initial step of soft BP decoding was implemented using the \texttt{bp decoder} in the \texttt{ldpc} package \cite{Roffe_LDPC_Python_tools_2022} with the product sum method and for $10d$ iterations for distance $d$ of the Chamon code.
This initial step improves  the final result significantly since it allows for global information to affect the decoding of each of the symmetry-subset MWPM decoders.

\begin{figure}
\includegraphics[scale=0.43]{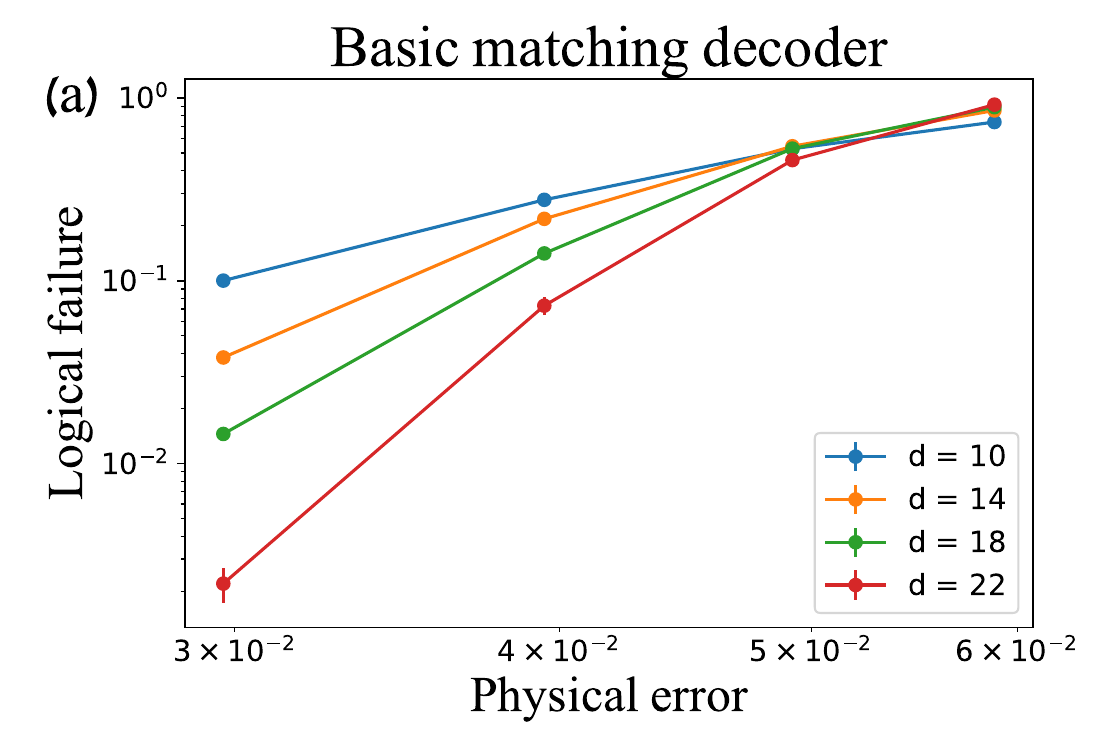}
\includegraphics[scale=0.41]{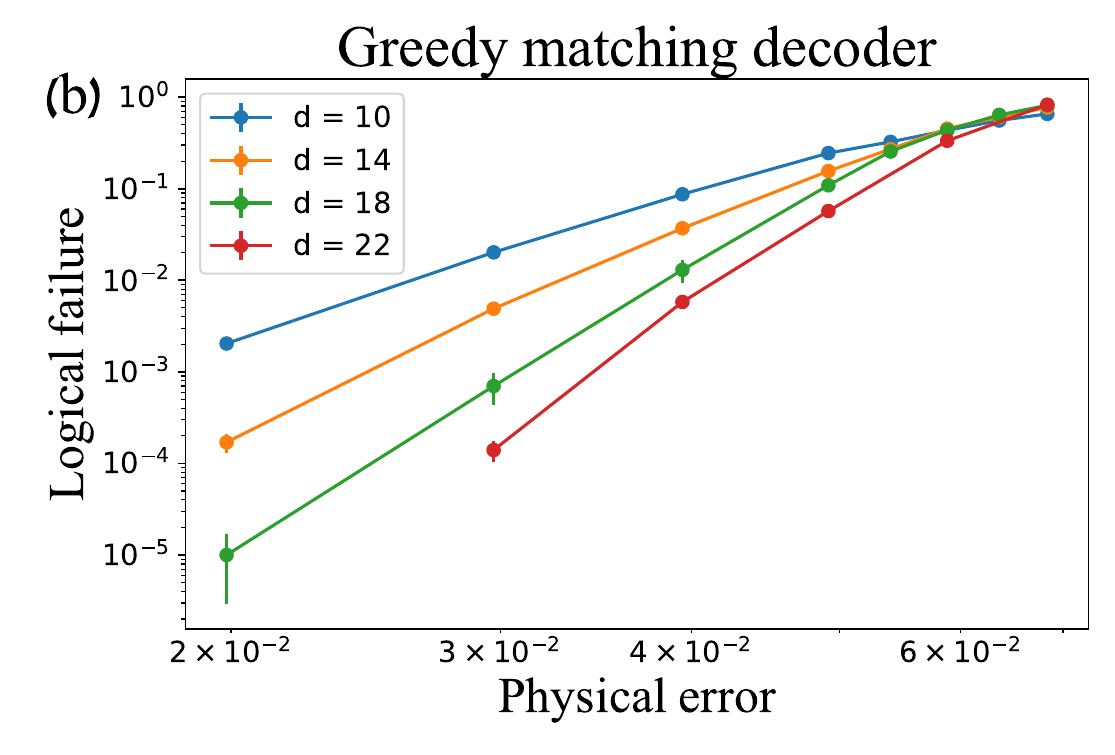}
\caption{\label{fig:logical failure naive} The total logical failure rate of any one of the considered logical qubits as a function of the physical error rate. The exponential decay of the failure rate for low errors, as well as the threshold can be observed. The standard deviation is presented in the error bars and was calculated by $\sqrt{\frac{P_{\text{fail}}-P_{\text{fail}}^{2}}{N}}$ where $P_{\text{fail}}$ is the logical failure rate and $N$ is the number of samples in the simulation.
(a) The standard matching decoder. The threshold can be seen to be $\sim 5\%$ .
(b) The greedy matching decoder.
For the improved matching decoder with an initial greedy step, the threshold is indeed slightly improved, and can be seen to be $\sim 6\%$. Also the exponential decay for low $p$ is significantly better, and logical error rates are significantly lower.}
\end{figure}

\section{Simulation and results}

The threshold of a QECC is one method to determine how well the code overcomes noise.
It is a property of a family of codes characterized by a parameter $L$. 
The threshold is the noise rate $p_*$ under which the logical error rate can be  suppressed arbitrarily by taking a representative code from the family with sufficiently large length $L$ \cite{knill1998resilient}.
This is not to be confused with another definition of threshold, sometimes also called a pseudo-threshold, which relates the logical error of a code to the physical noise \cite{aharonov1997fault}.

\begin{figure}
\includegraphics[scale=0.43]{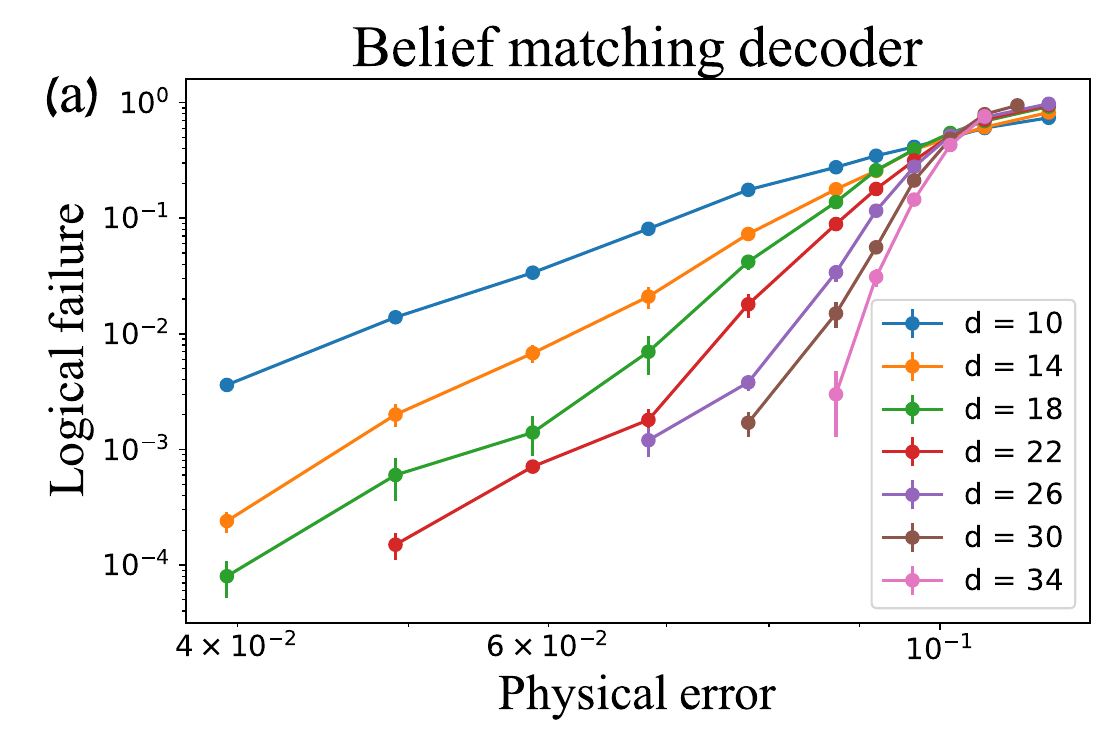}
\includegraphics[scale=0.44]{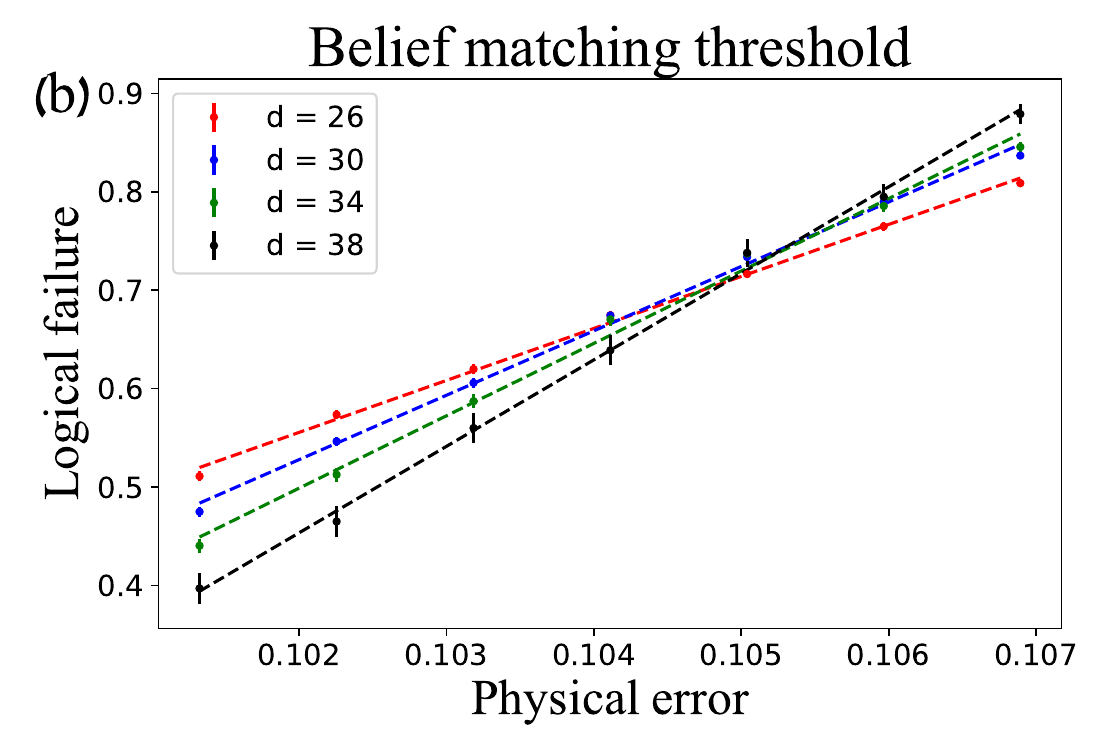}
\caption{\label{fig:logical failure with bp} The total logical failure rate of any one of the considered logical qubits as a function of the physical error rate for the Belief matching decoder. 
The standard deviation is presented in the error bars and was calculated by $\sqrt{\frac{P_{\text{fail}}-P_{\text{fail}}^{2}}{N}}$ where $P_{\text{fail}}$ is the logical failure rate and $N$ is the number of samples in the simulation.
(a) A large range of error rates shows the vast decay in logical failure rate for large system sizes. 
(b) A close-up on near threshold values with a linear fit of the data points reveals a threshold of $10.5\%$}
\end{figure}

The threshold is a function of the noise model we choose to analyse our system.
The noise model we consider here is the quantum depolarizing noise channel where all qubits of the system are affected identically by the map
\begin{equation}
    \mathcal{E}\left(\rho\right)=\left(1-p\right)\rho+\frac{p}{3}X\rho X+\frac{p}{3}Y\rho Y+\frac{p}{3}Z\rho Z
\end{equation}
where $p$ is the noise parameter that determines the strength of the noise.
This noise model is easy to analyze and assumes a symmetry between the different types of single qubit Pauli errors, and so is a natural choice of noise channel for assessing the logical error rate for the Chamon code.

We identify a logical failure rate when the noise together with the correction do not commute with any one of the eight logical operators presented in section 3.1 of \cite{zhao2023quantum}, which correspond to four logical qubits encoded in the code.
The total logical failure rate is the rate of failure of the sweep step of the decoder (which is negligible) added to the
logical failure rate.
The best threshold value attained in \cite{zhao2023quantum} for the Chamon code is $4.92\%$.
Already the basic implementation of generalized MWPM decoder slightly improves this threshold to a value of $\sim 5\%$, as can be seen in Fig. \ref{fig:logical failure naive}(a). 

\begin{figure}
\includegraphics[scale=0.45]{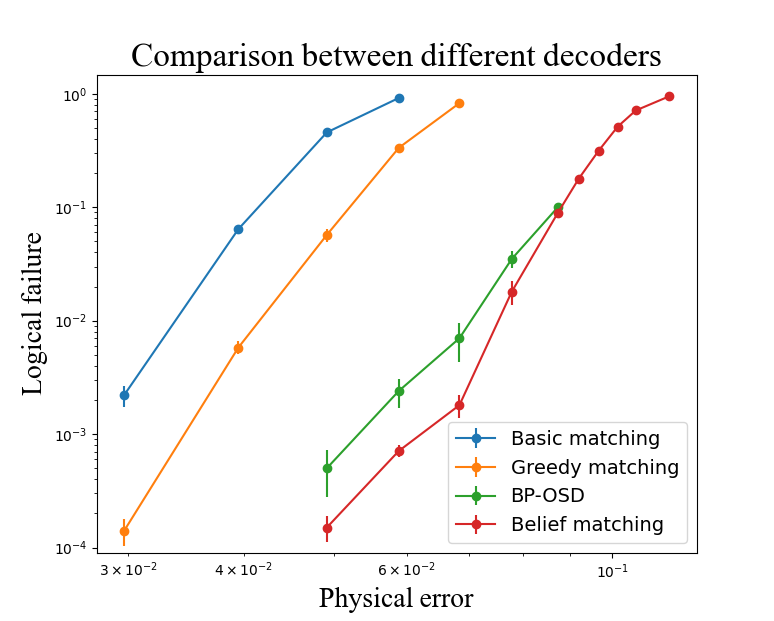}
\caption{\label{fig:dif decoders comparison} The total logical failure rate of any one of the considered logical qubits for the cubic $d=22$ Chamon code as a function of physical error rate $p$ for different decoders.
The belief matching decoder presented in this paper outperforms the other versions of matching decoders as well as the BP-OSD decoder.}
\end{figure}

Adding an initial greedy step further improves the threshold to $\sim 6\%$ as can be seen in Fig. \ref{fig:logical failure naive}(b). Notice that in \cite{anwar2014fast} a greedy initial step leads to improved decoding only for qudits with dimension 5 or larger, while for qubits and qutrits it leads to inferior decoding. In our case, though, due to the unique diamond shape of syndrome defects of a single-Pauli error, the greedy step leads to improved threshold even though we are dealing with qubits. 

An initial BP step has many advantages over an initial greedy step, and so we would expect it to lead to improved decoding. 
It makes a better estimate for the error based on soft decisions, instead of focusing on specific greedy patterns. We expect this to make the initial BP step less prone to mistakes compared with the greedy initial step. 
Indeed, the belief matching decoder obtains a threshold of $10.5\%$, as can be seen in Fig. \ref{fig:logical failure with bp}. This is a vast improvement compared to thresholds obtained with other decoders for the Chamon code.

Let us finally compare these decoders directly by concentrating on the logical error rates of a code of fixed distance, see Fig. ~\ref{fig:dif decoders comparison}. Here we show the performance of the decoders we have proposed together with a standard implementation of the belief-propagation ordered statistics decoder (BP-OSD)~\cite{Panteleev2021degeneratequantum, roffe_decoding_2020, bravyi2024high}. This decoder is well studied in the literature due to the fact that it is readily applicable to any code where the check matrix is known. We find that below threshold the belief matching on symmetries decoder performs best of all decoders.
We will also note that its runtime was significantly faster than BP-OSD. 
The complexity of the BP-OSD decoder is $O(n^3)$ \cite{Panteleev2021degeneratequantum}. 
On the other hand, the worst-scaling part of the belief matching decoder is the MWPM subroutine which scales as $\tilde{O}(n^3)$ \cite{higgott2022pymatching}. We run this subroutine on symmetries which include $O(n^{2/3})$ stabilizers and have $O(n^{1/3})$ such symmetries, and so in total the complexity of the belief matching decoder is $\tilde{O}(n^{7/3})$.
Given that both decoders begin using a belief-propagation step, our results indicate that matching subroutines may provide a better option than ordered statistics decoding when interpreting the results of belief propagation for cases where the message passing step does not converge.

\section{Summary}

We presented and analyzed a generalized MWPM decoder by exploiting the symmetries of the Chamon code, allowing for this code to enjoy the benefits of a matching decoder despite the non trivial structure of its syndrome data.
We also analyze two versions of improvements on this matching decoder by introducing initial decoding steps in order to incorporate the global data into the syndrome as well.
The best performance was achieved by the belief matching decoder where matching subroutines are informed by messages passed during an initial BP step.
This decoder has the best threshold between previously known decoders \cite{zhao2023quantum} and other variations of matching decoders presented in this paper. It also demonstrates the best below-threshold logical failure rates. This work demonstrates the potential in generalizing the belief-matching decoder to other QECCs with a non-trivial syndrome structure.
These results suggest that generalizing belief matching decoders to quantum LDPC codes~\cite{Panteleev2021degeneratequantum, Grospellier2021combininghardsoft, bravyi2024high} may help to access their full error-correcting potential, and this is an interesting avenue for future work.

A central challenge in this work, and the source of its most compelling open question, lies in the difficulty of identifying suitable symmetry planes on which the matching procedure can be applied.
These symmetry planes must meet two key criteria: (1) the matching decoder must be implementable on each symmetry plane, and (2) the clusters produced by the decoding processes on the symmetry planes must be independently correctable. 
Because of these constraints, identifying appropriate symmetry planes is a highly non-trivial task and, in this work, was carried out manually. This raises an important open question: is it possible to develop a general algorithm or systematic method for identifying such symmetry planes for arbitrary codes?

\begin{acknowledgements}
We thank D. Williamson for encouraging discussions, and J. Roffe for helpful explanations about BP decoding and their implementations. We also thank K. Sahay for comments on an earlier draft of the manuscript. BJB is grateful to the Center of Quantum Devices at the University of Copenhagen for their hospitality.
\end{acknowledgements}

\bibliographystyle{unsrtnat}
\bibliography{citations}

\end{document}